\documentclass{article}
\usepackage{spconf,amsmath,graphicx}
\usepackage{enumitem}
\usepackage{hyperref}
\usepackage{url}

\usepackage{cite}
\title{Singing-Tacotron: Global duration control attention and dynamic filter for End-to-end singing voice synthesis}
\name{Tao Wang$^{1,2}$, Ruibo Fu$^{1,2}$, Jiangyan Yi$^{1}$, Jianhua Tao$^{1,2,3}$, Zhengqi Wen$^{1}$}
\address{$^1$NLPR, Institute of Automation, Chinese Academy of Sciences, Beijing, China\\
 $^2$School of Artificial Intelligence, University of Chinese Academy of Sciences, Beijing, China\\
$^3$CAS Center for Excellence in Brain Science and Intelligence Technology, Beijing, China\\
}

\begin{document}

\ninept
\maketitle

\begin{abstract}
End-to-end singing voice synthesis (SVS) is attractive due to the avoidance of pre-aligned data. However, the auto learned alignment of singing voice with lyrics is difficult to match the duration information in musical score, which will lead to the model instability or even failure to synthesize voice. To learn accurate alignment information automatically, this paper proposes an end-to-end SVS framework, named Singing-Tacotron. The main difference between the proposed framework and Tacotron is that the speech can be controlled significantly by the musical score’s duration information. Firstly, we propose a global duration control attention mechanism for the SVS model. The attention mechanism can control each phoneme’s duration. Secondly, a duration encoder is proposed to learn a set of global transition tokens from the musical score. These transition tokens can help the attention mechanism decide whether moving to the next phoneme or staying at each decoding step. Thirdly, to further improve the model’s stability, a dynamic filter is designed to help the model overcome noise interference and pay more attention to local context information. Subjective and objective evaluation \footnote{Examples of experiments can be found at \href{https://hairuo55.github.io/SingingTacotron}{https://hairuo55.github.io/SingingTacotron.}} verify the effectiveness of the method. Furthermore, the role of global transition tokens and the effect of duration control are explored.

\end{abstract}

\section{INTRODUCTION}

\label{sec:intro}
Due to the powerful modeling capabilities of deep neural networks (DNN) \cite{goodfellow2016deep}, different kinds of DNN-based  singing voice synthesis (SVS) models \cite{ling2015deep,hono2018recent,nishimura2016singing,nakamura2019singing} are proposed which can produce more natural acoustic features than traditional hidden Markov model (HMM)-based SVS   \cite{saino2006hmm,8659797,tokuda2013speech,black2007statistical}.
 In a typical DNN-based SVS model, a DNN works as an acoustic model representing a mapping function from the musical score  sequence to the acoustic feature sequence. Since the input and output feature needs to be time-aligned frame-by-frame, such models still have some limitations.

\begin{figure}[tp]
 \centering
 \begin{minipage}[t]{0.4\textwidth}
  \centering
  \centerline{\includegraphics[width=7.5cm]{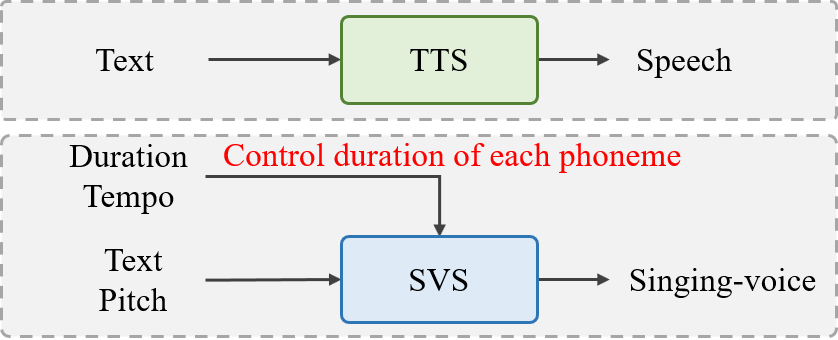}}
\end{minipage}
\vspace{-0.1cm}
\caption{Different from TTS system, the time information (duration and tempo) of musical score in SVS system significantly controls the pronunciation time of each phoneme.}
\label{fig:different}
\vspace{-0.5cm}
\end{figure}
One limitation is that the system needs a pre-trained duration model to expand the musical score feature sequence \cite{gu2021bytesing,lu2020xiaoicesing}. Some tools, such as force alignment \cite{black2007statistical,tokuda2013speech} with an HMM model, can help us get the alignment information of phonemes and acoustic features. However,  the duration model can not yield sufficiently accurate results on expressive singing, requiring manual correction and costing a lot of time. Another limitation is that it is easy for the pipeline model to accumulate errors and increase system construction difficulty.

To overcome the above problems, some end-to-end SVS models are proposed \cite{blaauw2020sequencetosequence,angelini2020singing,lee2019adversarially}.  Like Tacotron in the task of text-to-speech (TTS), the end-to-end SVS model is to input musical score and output acoustic features. However, different from the TTS model, as shown in Fig. \ref{fig:different}, there is extra duration information in the music score, which requires effective modeling to achieve better results. So the main idea is to combine the duration information as a constraint during training \cite{blaauw2020sequencetosequence,angelini2020singing}. Adversarial training is also adopted to improve the accuracy of predicted features \cite{lee2019adversarially}. However, since the attention mechanism is generally based on the content and does not reflect the duration constraint, the end-to-end model usually suffers from a lack of robustness in alignment. Therefore, if the attention mechanism can  be controlled by the duration information in the musical score, the SVS system will be more stable and controllable.

This paper proposes an end-to-end SVS model to improve the quality and stability of synthesized voice through global duration control attention and dynamic filter. Firstly, to make the attention mechanism controlled by duration information, a set of global transition tokens is learned from the musical score to control each phoneme's duration more precisely. Secondly, to improve the robustness of the end-to-end model, we design a way of dynamic filtering in decoding, making the model only focus on the local context information. In summary, the main contributions are as follows:
\vspace{-0.2cm}
\begin{itemize}
\item  We propose a global transition control attention for the end-to-end SVS model, which uses a set of learned global transition tokens to help the attention control the speed of decoding. Experiments show this attention can be controlled by duration information effectively and learn accurate alignment.
\item A dynamic filter is set up for each decoding step to overcoming noise interference on the model. Experiments show that it can improve synthesis speech's quality.
\end{itemize}

The rest of the paper is organized as follows. Section \ref{2} describes the methods. Experiments and results are analyzed in Section \ref{3}. The conclusions are discussed in Section \ref{4}.

\begin{figure*}[t]
 \centering
 \begin{minipage}[t]{0.4\textwidth}
  \centering
  \centerline{\includegraphics[width=17.5cm]{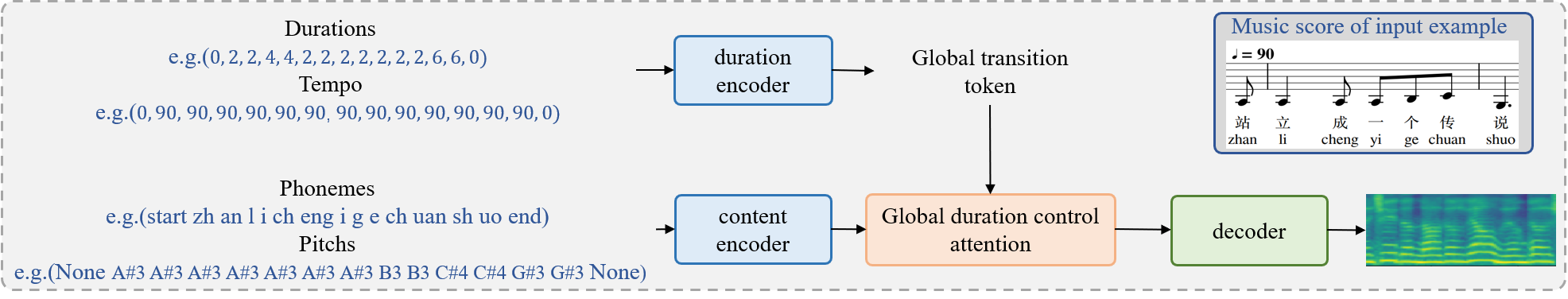}}

 \end{minipage}
\vspace{-0.3cm}
 \caption{The architecture  of our proposed end-to-end SVS model: Singing-Tacotron.}
 \label{fig:architeture}
\vspace{-0.4cm}
\end{figure*}



\vspace{-0.3cm}

\section{proposed method}\label{2}
\vspace{-0.2cm}
In the application of SVS, a notable difference from TTS is that the alignment between the input phoneme and output acoustic feature is strongly constrained by the duration information in the  musical scores.  To enable the attention mechanism to be controlled by duration information, this section will introduce the whole framework first,  then we will present the ideas of global duration control attention and dynamic filter in detail.
\vspace{-0.3cm}

\label{sec:pagestyle}
\subsection{The framework of Singing-Tacotron}
\vspace{-0.1cm}
Our proposed framework Singing-Tacotron is shown in the Fig. \ref{fig:architeture}. The whole framework consists of four parts: the content encoder is mainly responsible for encoding lyrics information into hidden features. The input of this module is a musical score consisting of a sequence of notes. Each note consists of an onset,  a sequence of phonemes and pitches, typically corresponding to a syllable. The duration encoder encodes the parameters related to the duration information (the duration and the tempo in musical score) into a global transition token, which is used to help the attention decide whether to move to the next phoneme not. The global duration control attention connects the two encoders  and the decoder, and controls each phoneme's duration according to the global transition token. The decoder generates acoustic features, which can be restored to the singing voice by vocoder \cite{valin2019lpcnet}. 
\vspace{-0.4cm}

\subsection{Global duration control attention}
\vspace{-0.1cm}
To enable the duration information to play an influential role in the attention mechanism, we propose a global duration control attention.  There are two assumptions in the attention. 

\begin{figure}[tp]
 \centering
 \begin{minipage}[t]{0.4\textwidth}
  \centering
  \centerline{\includegraphics[width=6cm]{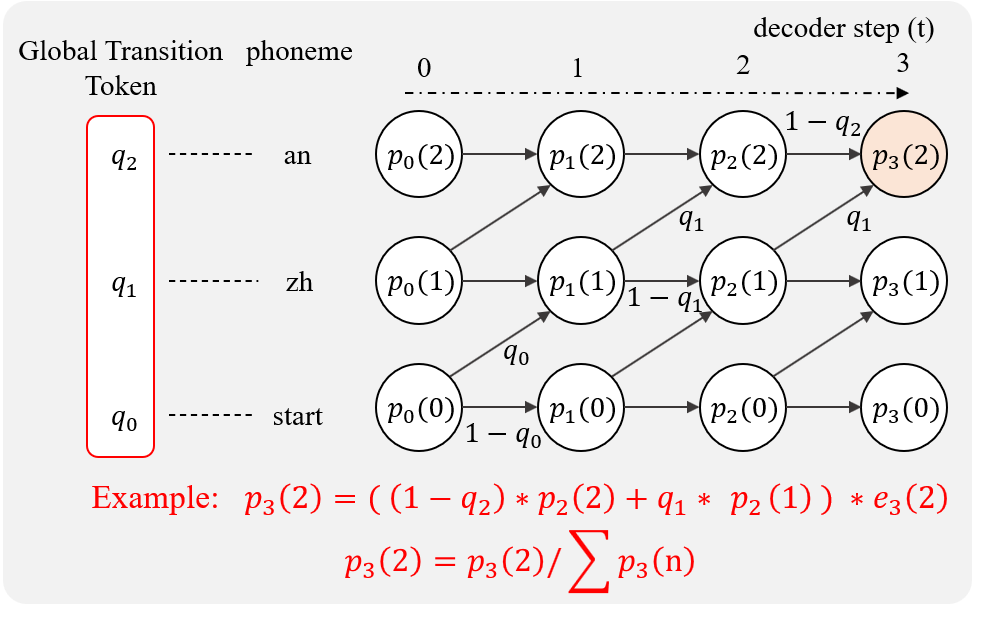}}
\end{minipage}
\vspace{-0.2cm}
\caption{Global duration control attention. There are only three phonemes as input for easy presentation.  $p_{t} (n)$ represents  the probability that the attention mechanism notices the n-th phoneme  at the t-th decoder step.  $q_{n}$ represents the transition probability of the n-th phoneme, which is controlled by the duration information in the musical score. $e_{t} (n)$ represents the soft alignment calculated based on the lyrics content. $p_{t} (n)$ can be calculated recursively.}
\label{fig:ali}
\vspace{-0.5cm}
\end{figure}

Firstly, inspired by forward attention \cite{8462020}, to meet the monotonic alignment between speech and phonemes, we assume that the alignment path moves monotonically and continuously without skipping any encoder states. Specifically, the phoneme noticed by the attention mechanism in the current decoding step must be the next phoneme or the same phoneme of the phoneme noticed in the previous decoding step.  For example, suppose the probability that the attention mechanism notices the n-th phoneme  at the t-th decoder step is $p_{t}(n) $ , then $p_{t}(n) $ is related to $p_{t-1}(n) $ and $p_{t-1}(n-1)$ in the previous step.  As shown in Fig .\ref{fig:ali}, the value of $p_{3}(2) $ comes from the accumulation of $p_{2}(2) $ and $p_{2}(1) $.  To ensure that the alignment starts with the first phoneme, we initialize $p_{0}(0) =1 $ and $p_{0}(i) = 0, i =1, \cdots, N-1 $. Where N is the number of phonemes.

Secondly, in the SVS task, each phoneme's pronunciation time should be also controlled by the given duration information. Therefore, another assumption is that 
in two adjacent decoding steps, the probability of attention mechanism shifting from the n-th phoneme to the (n+1)-th phoneme is $q_n$, and the $q_n$ is only controlled  by the  duration information in musical score. This assumption is inspired by the fact that the lyrics information and the duration information are independent of each other in the musical score. For example, As shown in Fig .\ref{fig:ali},  the probability of $p_{1}(1)$ transferring to $p_{2}(2)$ is $q_1$.
In addition, according to the first assumption, the alignment path moves continuously and can not skip, so the probability of n-th phoneme keeping unmoved is $1-q_n$.  As shown in Fig .\ref{fig:ali},  the probability of $p_{1}(1)$ transferring to $p_{2}(1)$ is $1-q_1$.

Based on the above two assumptions, given an input sequence $\boldsymbol{x} = [\boldsymbol{x}_0, \boldsymbol{x}_2, \cdots, \boldsymbol{x}_{N-1}]$ with length $N$, which is a combined sequence of phoneme and pitch information. The content encoder first processes $\boldsymbol{x}$ as a sequence of hidden represents $\boldsymbol{h} = [\boldsymbol{h}_0, \boldsymbol{h}_2, \cdots, \boldsymbol{h}_{N-1}]$.  By inputting the duration sequence and tempo sequence, the global transition token $\boldsymbol{q} = [q_0,q_2, \cdots, q_{N-1}]$ is also predicted by the duration encoder.
Then the decoder generates each output $\boldsymbol{O}_t$ conditioned on the hidden represents $\boldsymbol{h}$ and global transition token $\boldsymbol{q}$.  The process is as follows.

At each decoder timestep $t$, let $\boldsymbol{m}_t$ denote the query of the output sequence at the t-th decoder step which is usually the hidden state of the decoder.  A content-based attention mechanism \cite{shen2018natural} is first used to calculate soft alignment information $e_{t}(n)$ based on content information $\boldsymbol{h}$, which can be expressed  as:
\vspace{-0.1cm}
\begin{align}
e_{t}(n)&=\boldsymbol{v}^{T} \tanh \left(\boldsymbol{W} \boldsymbol{m}_{t}+\boldsymbol{V} \boldsymbol{h}_{n}+\boldsymbol{b}\right) \\
e_{t}(n)&=\exp \left(e_{t}(n)\right) / \sum_{N} \exp \left(e_{t}(n)\right)
\vspace{-0.2cm}
\end{align}

Second,  according to the above two assumptions, $p _t(n)$ can be calculated recursively from $p _{t-1}(n)$ and $p_{t-1}(n-1)$ as:
\begin{align}
p _t(n) = (1-q _{n-1}) \cdot p _{t-1}(n-1) + q _{n} \cdot p _{t-1}(n)   \label{eqa}
\end{align}
Eq. \ref{eqa} contains duration information (included in $q _{n}$ and $q _{n-1}$), but does not contain content information, which contains in $e_{t}(n)$. Since the content and duration information are independent of each other, in order to combine the two, we recalculate $p _t(n)$ by multiply $e_{t}(n)$ and $p _t(n)$ to control the decoding process at the same time:
\begin{align}
p _t(n) = ((1-q _{n-1}) \cdot p _{t-1}(n-1) + q _{n} \cdot p _{t-1}(n) ) \cdot e_{t}(n) \label{eq0}
\end{align}
Then, to make sure the sum of $p _t(n)$ for the t-th timestep to be 1 and substitute $p _t(n)$ for $e_{t}(n)$ to calculate the context vector,  $\boldsymbol{p}_{t}$ is  normalized and the context vector $\boldsymbol{c}_{t}$ derived from the input is calculated as:
\vspace{-0.2cm}
\begin{align}
p_{t}(n) &= p_{t}(n) / \sum_{N} p_{t}(n) \label{eq1} \\
\boldsymbol{c}_{t}&=\sum_{N} {p}_{t}(n) \boldsymbol{h}_{n}
\vspace{-0.3cm}
\end{align}

 Finally, the output vector $\boldsymbol{O}_t$ can be computed conditioning on the context $\boldsymbol{c}_{t}$.

Additionally, we can change the Eq. \ref{eq0} to get:
\begin{align}
p _t(n) = (1-q _{n-1} , q _{n} )   \cdot \begin{pmatrix} p _{t-1}(n-1) \\ p _{t-1}(n) \end{pmatrix}  \cdot e_{t}(n) \label{eq2}
\end{align}
Obviously, $ (1 - q _{n-1} , q _{n} )$  is controlled by the duration information in the musical score, and $e_{t}(n)$ is decided by the lyric information. $ (1 - q _{n-1} , q _{n} )$ is independent of time $t$. This means that the $q$ can be used as a global parameter to guide the transfer of the alignment path during the entire decoding process. Therefore, we name the $ \boldsymbol{q}$ as global transition token. 
\vspace{-0.2cm}

\subsection{Dynamic filter in decoding}
Through the global duration control attention, we can integrate the lyrics content and duration information into the decoding process, which is shown in Eq. \ref{eq2}. In Eq. \ref{eq2},  $e_t(n)$ is the soft alignment information calculated according to the content-based attention.  However, the content-based attention is easily affected by noise. For example, the attention may produce greater attention to an irrelevant text \cite{chorowski2015attention}, which will lead to difficulties in model training.

To improve the robustness of the model, inspired by the attention mechanism in speech recognition task \cite{chorowski2015attention}, we propose a method of dynamic filtering in the decoding process. We assume that in each step of decoding, only some partial information of encoder states is needed. Firstly, we find the maximum value of $\boldsymbol{p} _{t-1}$ in Eq. \ref{eq1} and record the index $m$ corresponding to the maximum value. This means that the current decoding mainly focuses on the $m$-th phoneme. Then, we add a window function of width $L$ with $m$-th phoneme as the center.  The numbers outside the window are masked with zero. For example, when we use the rectangular window as a window function, and the index of the maximum value of $\boldsymbol{p} _{t-1}$ is $m$. After filtering by the window function, a new $p _{t-1}{'}$ is obtained as: $[0,\cdots,p _{t-1}(m-L/2),\cdots,p _{t-1}(m),\cdots,p _{t-1}(m+L/2),\cdots,0]$.  Then we bring $\boldsymbol{p} _{t-1}{'}$ into the formula Eq. \ref{eq0} and Eq. \ref{eq1}  to compute the new alignment.
\vspace{-0.2cm}

\begin{figure}[t]
 \centering
 \begin{minipage}[t]{0.4\textwidth}
  \centering
  \centerline{\includegraphics[width=7cm]{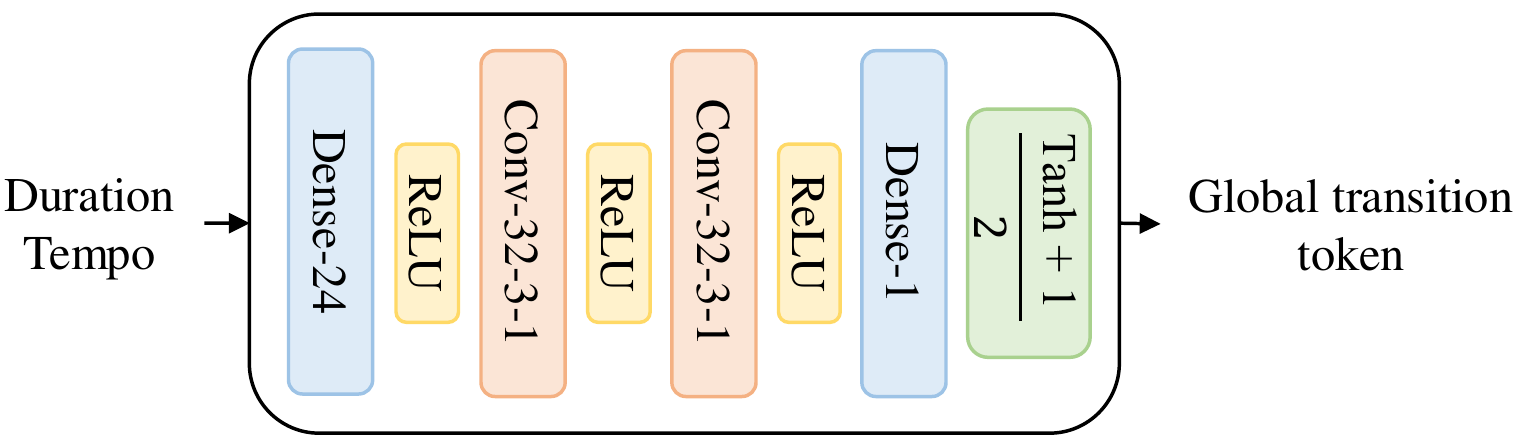}}
\end{minipage}
\vspace{-0.2cm}
\caption{The structure of duration encoder.}
\label{fig:gtt}
\vspace{-0.6cm}
\end{figure}

\section{Experiments}\label{3}
\label{sec:typestyle}
To verify the validity of the method, since there are few open source data sets about singing voice, 100 Chinese songs \cite{biaobei} performed by a female singer are used. We use 95 songs as the training dataset, and another 5 songs are used as the test dataset to measure the performances. The musical scores and lyrics are annotated on MusicXML format \cite{Good2006MusicXMLIC}. All the wav files are sampled at 16KHz.

\subsection{Setup}
In our proposed model, the duration encoder's framework is shown in Fig. \ref{fig:gtt}, and the purpose of the last layer is to make the output value between $[0,1]$. The content encoder and the decoder module are the same as the encoder and decoder in tacotron2 \cite{shen2018natural}. Acoustic features are extracted with a 10 ms window shift.
LPCNet \cite{valin2019lpcnet} is utilized to extract 32-dimensional acoustic features.
To evaluate the effects of global duration control attention and the dynamic filter, the following six systems are established for comparison.
\begin{itemize}
\item Local sensitive attention, without or with window technique \cite{chorowski2015attention}, which are denoted as \textbf{LA} and \textbf{LA+Window}. The window technique in this attention is similar to the dynamic filter in global duration control attention. $e_t(n)$ is directly used as alignment information in Eq. \ref{eq0}.
\item Forward attention, without or with dynamic filter setting, denoted as \textbf{FA} and \textbf{FA + DF}. That is, $1-q_{n-1}$  and  $q_{n}$ are removed in Eq. \ref{eq0}.
\item Global duration control attention, without or with dynamic filter setting, which are denoted as \textbf{GDCA} and \textbf{GDCA + DF}. 
\item \textbf{ByteSing}\cite{gu2021bytesing}  with duration model based on LPCNet vocoder.
\end{itemize}
\textbf{It should be noted} that since the models based on location sensitive attention and forward attention do not have a duration encoder, for the fairness of comparison, we link the duration and tempo information to the content encoder of these models to ensure the integrity of the information. The width of all windows in the dynamic filter or window technique is set to 16.
\vspace{-0.3cm}

\subsection{The stability of alignment}
\begin{figure}[t]
\centering
\begin{minipage}[t]{0.4\textwidth}
 \centering
 \centerline{\includegraphics[width=9cm]{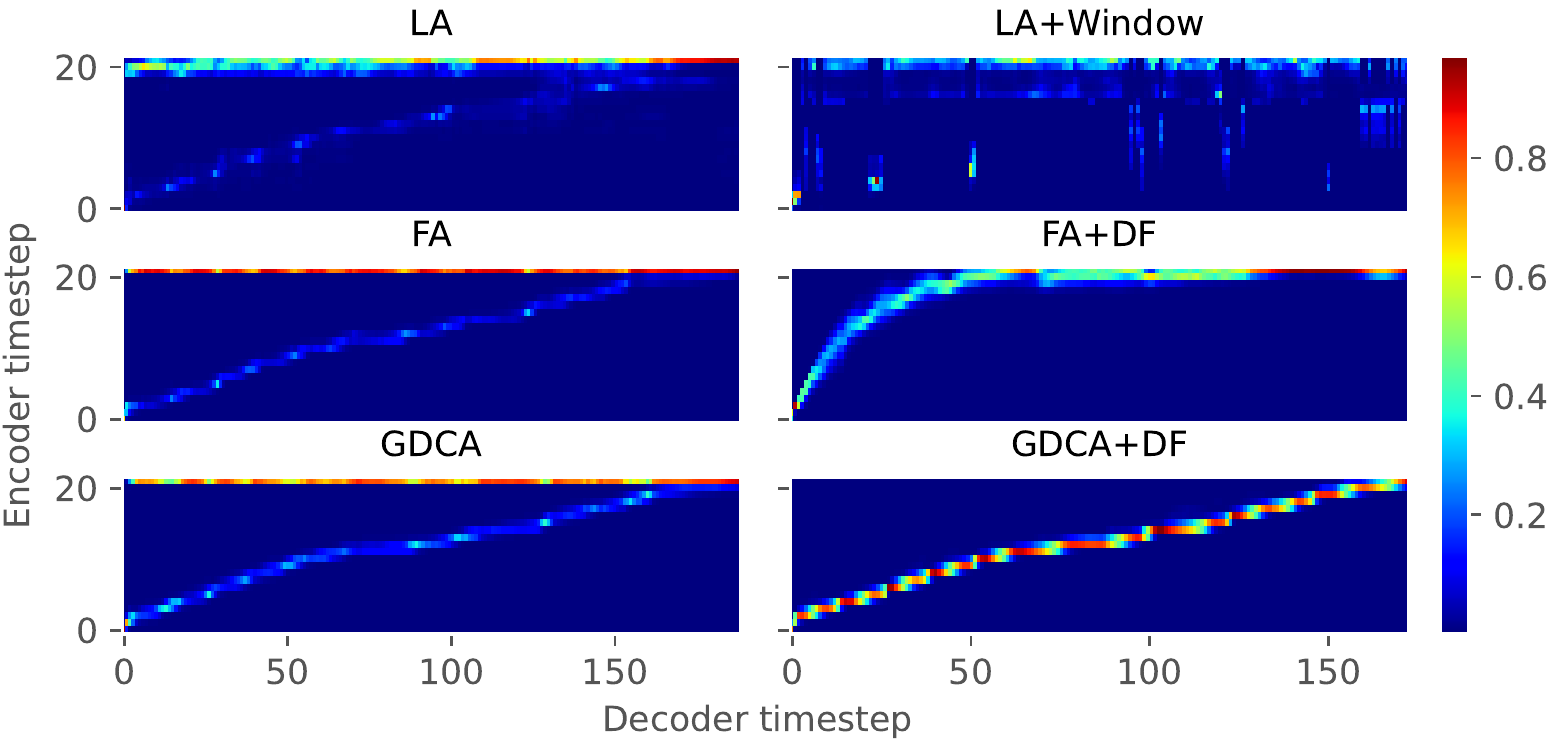}}
\end{minipage}
\vspace{-0.2cm}
\caption{Alignments of a lyric given by different systems}
\vspace{-0.5cm}
\label{fig:individualImage}
\end{figure}

First, we use the alignment diagram between phonemes and acoustic features to illustrate the effect of models through ablation experiments, which is shown in Fig. \ref{fig:individualImage}. First, by comparing the three attention mechanisms, we find that the monotonic intensity relationship of the alignment information is $\textbf{GDCA}>\textbf{FA}>\textbf{LA}$. Among them, because \textbf{LA} does not have any prior knowledge, it is hard to learn effective alignment information. \textbf{FA} has more obvious alignment than \textbf{LA} because of monotonic prior knowledge, and \textbf{GDCA} has not only monotonic prior knowledge but also can be controlled by duration information, so as to be able to learn the most apparent alignment information.
Second, by comparing a dynamic filter setting, we find that adding the dynamic filter can help the global duration control attention obtain more stable and monotonous alignment information, corresponding to model \textbf{GDCA+DF}. For forward attention, adding dynamic filtering (\textbf{LA+DF}) causes model synthesis to fail. This also shows that the assumption of monotonicity alone is not enough to make the SVS model robust, and the assumption of duration control is also required.

\subsection{Objective  and subjective evaluation}
\begin{table}[t]
    \centering
        \caption{Objective evaluation results of each system.}
\scalebox{0.85}{
\begin{tabular}{c|ccccc}
\hline \hline     & FA      & FA+DF   & GDCA  & ByteSing  & \textbf{GDCA+DF}  \\
\hline   MCD(dB) $\downarrow$  & 3.204   & 4.302   &3.073 & 3.119   & \textbf{3.055}\\
  F0-RMSE(Hz)  $\downarrow$   & 27.497  & 43.343  &19.374 & 19.294 & \textbf{17.884}\\
   V/UV error(\%) $\downarrow$ & 3.940   & 7.985   &3.743 & 4.013  & \textbf{3.712} \\ 
 F0-CORR     $\uparrow$      &0.884    & 0.682   &0.944   & 0.948 & \textbf{0.953}\\

\hline \hline
\end{tabular}}
\label{table:metric}
\vspace{-0.5cm}
\end{table}

\begin{table}[t]
  \caption{Mean Opinion Score (MOS) ratings on a 1-5 scale with
their respective 95\% confidence intervals.}
  \centering
  \begin{tabular}{ c | c  }
    \hline \hline
    {\textbf{Model}} &
                  {\textbf{MOS}}  \\
    \hline
    \text{LA  \& LA+Window  \& FA+DF}   & $\text{Failed}$ ~~~  \\
    FA & $3.80 \pm 0.12$  ~~~ \\
    GDCA  & $4.02  \pm 0.08$ ~~~  \\
ByteSing & $3.97  \pm 0.06$ ~~~  \\
    \textbf{GDCA+DF} & $\textbf{4.14}  \pm \textbf{0.08}$ ~~~ \\
    Recording & $4.58 \pm 0.08 $ ~~~\\
    \hline \hline
  \end{tabular}
  \label{fig:mos}
\vspace{-0.5cm}
  \end{table}

First, F0 RMSE (root of mean square errors of F0), MCD (Mel-cepstrum distortion), VUV (the error rate of voiced/unvoiced flags), and F0 CORR(correlation factor of F0) were adopted as metrics and were calculated on the test sets. For two different lengths of
speech, we  use the DTW algorithm to align them. The objective results  are listed in Table \ref{table:metric}.  In general, it can be found that the metrics  of \textbf{GDCA+DF} are the best among all the systems. Specifically, in the frequency domain, the \textbf{GDCA+DF} obtained the lowest MCD, which means that human perception would be better. Besides,  \textbf{GDCA+DF} achieves the best performance in F0-related metrics (F0-RMSE, V/UV error,  F0-CORR). The results show that our method can obtain more accurate F0 information. 

Second, we conduct the Mean Opinion Score (MOS) listening test for audio quality on the test set. Twenty listeners  take part in the evaluation.  They were asked to listen and rate the quality of the synthesized singing song  on a scale from 1 to 5. We keep the lyrics consistent among different models to examine the audio quality. Fifty sentences are randomly chosen from the test dataset and used for the evaluation. The results are shown in Table \ref{fig:mos}. Among all systems, \textbf{GDCA+DF} achieves the best performance. Additionally, we can find that the global duration control attention plays an essential role in the end-to-end SVS model, ensuring the model synthesizes a stable and natural singing voice. Then the dynamic filter in decoding can further improve the quality of the synthesized

\vspace{-0.2cm}

\begin{figure}[t]
 \centering
 \begin{minipage}[t]{0.4\textwidth}
  \centering
  \centerline{\includegraphics[width=8cm]{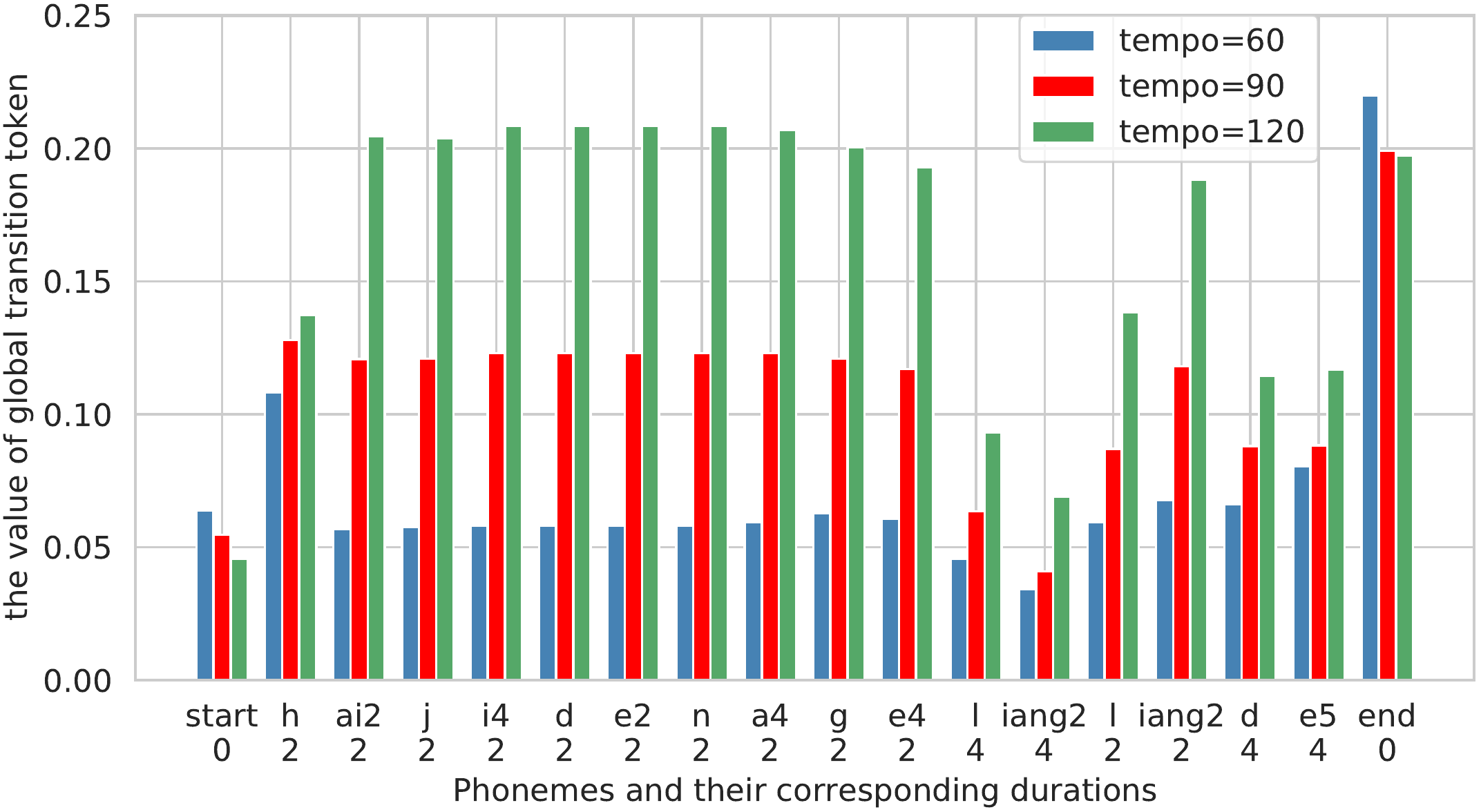}}
 \end{minipage}
\vspace{-0.2cm}
 \caption{The value of global transition token in a lyric with different tempos.}
 \label{fig:gtt_value}
\vspace{-0.2cm}
\end{figure}
\subsection{Visualization of global transition tokens}
To explain the role of the learned global transition token to the network, we visualize the token values in lyrics with different tempos, shown in Fig. \ref{fig:gtt_value}. Firstly, we observe the token values corresponding to each phoneme with the same tempo value. We can find that because each phoneme value corresponds to a different duration, the corresponding token values are different. Additionally, the larger the duration value, the smaller the token value, which shows that the lower the phoneme's probability of moving to the next one. Secondly, when the same phoneme corresponds to different tempo values, we can find that the larger the tempo value, the longer the phoneme duration, that is, the larger the token value. In summary, we can conclude that the token value can be controlled by duration and tempo information.
\vspace{-0.2cm}



\subsection{Duration control with different tempos and durations}
\begin{figure}[t]
 \centering
 \begin{minipage}[t]{0.4\textwidth}
  \centering
  \centerline{\includegraphics[width=9.1cm]{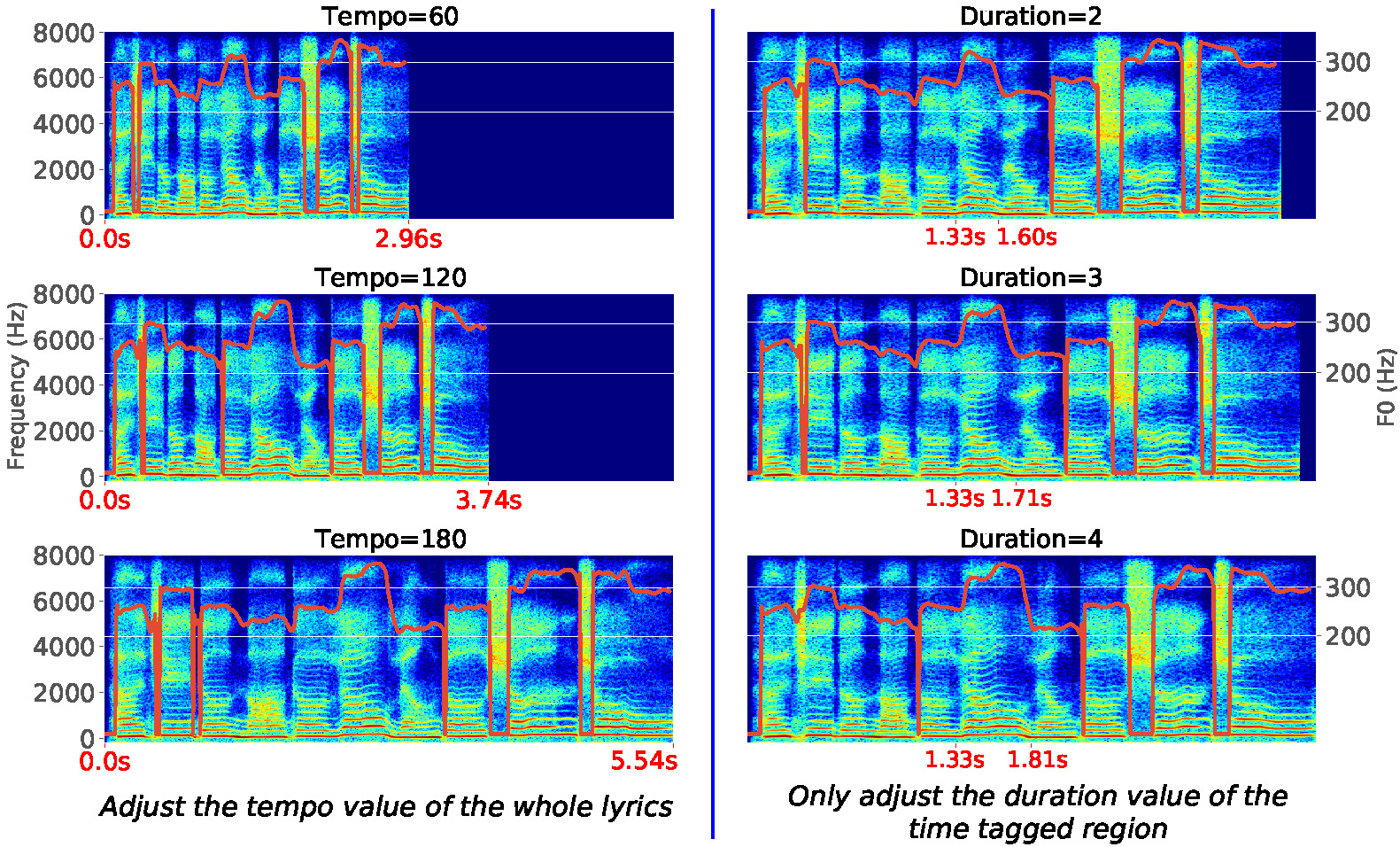}}
 \end{minipage}
\vspace{-0.2cm}
 \caption{Mel-spectrograms with different tempos (left part) and durations (right part).}
 \label{fig:mel}
\vspace{-0.4cm}
\end{figure}


According to the global transition token analysis in the previous section, we find that the speed of singing voice can be controlled by adjusting the tempo values and duration values in the musical scores. 

First, we set different tempo values ($60$, $120$ and $180$) for the whole lyrics. As shown in left part of  Fig. \ref{fig:mel}, our model generates speech with different lengths. The results show that the model can control the velocity of the speech without quality degradation.

Second, we only adjust a single word's duration information (the time tagged area)  in lyrics, as shown in the right part of Fig. \ref{fig:mel}. It can be found that the duration of synthesized speech changes in a similar positive proportion with the change of the duration information of the word. This phenomenon shows that this method can accurately control the pronunciation duration of synthesized speech.

In summary, the proposed model can effectively control the duration of the whole speech or the duration of a single word through the tempo value or duration value in the music score.



\vspace{-0.2cm}
\section{Conclusion}\label{4}
\label{sec:prior}
This paper presents an end-to-end SVS model with a global duration control attention mechanism, aiming to make the attention mechanism controlled by duration information and improve the naturalness of synthesis. First, by incorporating the monotonicity and duration-controlled characteristics of the alignment path, a set of global transition tokens is learned from duration information to obtain more accurate alignment information. Second,  to improve the  robustness, a dynamic filter is set up for each decoding step to help the model pay more attention to useful information. Experiments demonstrate that the proposed model can learn the time constraint information effectively and synthesize natural voice.  Further, we will combine more musical theories with improving the synthesis effect.

\vfill\pagebreak

\bibliographystyle{IEEEbib}
\bibliography{strings,refs}

\end{document}